# Examining the sentiment and emotional differences in product and service reviews: The moderating role of culture

Dr Vinh Truong, RMIT University, vinh.truongnguyenxuan@rmit.edu.vn

## Abstract

This study investigates the emotional and sentiment differences in customer reviews of products and services on e-commerce platforms. Unlike prior research that treats customer reviews as a single entity, this study distinguishes between products, which fulfil fundamental needs, and services, which cater to higher-level experiential desires. The analysis reveals that customer sentiment and emotional expressions vary significantly between these two categories, reflecting their distinct roles in consumer decision-making. One of the key findings is that products tend to receive reviews centred around functionality, reliability, and value for money, often characterized by a more neutral or pragmatic tone. In contrast, services evoke stronger emotional responses, as they involve direct interactions, subjective experiences, and personal satisfaction. Customers express a wider range of emotions, including joy, frustration, and disappointment, when reviewing services compared to products as detected by recently developed machine learning techniques. Cultural differences further amplify these distinctions. Consumers from collectivist cultures, as identified by Hofstede's cultural dimensions, tend to emphasize group consensus and social harmony in their reviews, often using more moderated language. Meanwhile, those from individualist cultures are more likely to provide direct and emotionally intense feedback. Interestingly, gender does not significantly influence sentiment differences, suggesting that product and service categorization and cultural background are the primary drivers of variation in reviews. Theoretically, this study extends Maslow's hierarchy of needs and Hofstede's cultural framework by applying them to consumer reviews, offering a conceptual model of how these factors shape customer experiences. From a practical perspective, businesses can leverage these insights to refine their marketing and customer engagement strategies, tailoring their communication and service approaches to align with customer expectations based on product and service offerings and cultural context.

## 1. Introduction

The rapid expansion of e-commerce platforms has transformed consumer behaviour, influencing how individuals purchase and review products and services. Customer reviews serve as a vital source of information for both businesses and potential buyers, shaping purchasing decisions and brand perceptions (Chevalier & Mayzlin, 2006; Zhu & Zhang, 2010). Prior research has extensively examined the role of online reviews in driving sales, enhancing consumer trust, and predicting purchase behaviour (Filieri, 2015; Floyd et al., 2014; Mudambi & Schuff, 2010).

However, most studies have treated customer reviews as a homogeneous dataset, failing to differentiate between products and services despite their fundamental distinctions in consumer expectations and consumption experiences (Vargo & Lusch, 2004; Zeithaml, 1988). This oversight represents a critical gap in the literature, as the factors that shape customer reviews may differ significantly depending on whether the review pertains to a product, which fulfils tangible and functional needs, or a service, which caters to experiential and relational needs (Sudirjo et al., 2023).

Existing research on online reviews has primarily focused on their impact on consumer decision-making and brand reputation management (Filieri et al., 2018; You et al., 2015). Studies have analyzed review valence, volume, and credibility to understand their influence on purchase behaviour (Dellarocas et al., 2007; Hennig-Thurau et al., 2004). However, these studies have largely neglected the inherent differences in consumer expectations and emotional responses when reviewing products versus services. Given that products primarily fulfil physiological and safety needs, while services address higher-order psychological and social needs (Maslow, 1943), it is plausible that consumer sentiments and emotions in reviews would differ across these categories (Chiu et al., 2012). Despite this theoretical basis, empirical investigations into these differences remain scarce, leaving an important gap in the understanding of how consumer expectations shape review sentiment and emotional expression.

Another limitation in the literature is the narrow focus on sentiment polarity (positive, negative, neutral) rather than a nuanced analysis of emotions. While studies in sentiment analysis have explored basic emotions (joy, anger, sadness, etc.), few have examined emotions beyond valence and arousal (Guo et al., 2024; Pang & Lee, 2008). Recent advancements in psychological and computational emotion research suggest that a more comprehensive framework—including valence, arousal, and dominance (VAD) alongside advanced discrete emotional categories—provides a deeper understanding of consumer sentiment (Buechel & Hahn, 2017). Despite these advancements, limited research has applied this expanded emotional framework to customer reviews in e-commerce settings, particularly in differentiating between product and service evaluations (Truong et al., 2020). Addressing this gap, this study seeks to analyze a more detailed spectrum of emotional expressions across product and service reviews.

Cultural variations in consumer behaviour further complicate how sentiments and emotions are expressed in online reviews. Hofstede (2011)'s cultural dimensions theory suggests that individualism, uncertainty avoidance, and long-term orientation influence communication styles and consumer decision-making. Prior research indicates that collectivist cultures emphasize social harmony, leading to more restrained and indirect expressions of dissatisfaction, while individualistic cultures tend to provide explicit and detailed feedback (Park et al., 2007; Schuckert et al., 2015). Additionally, high-context cultures, which rely on implicit communication, may express emotions through subtle linguistic cues rather than direct sentiment words (De Kock et al., 2018). While these cultural differences have been studied in customer feedback contexts (Liu et al., 2023), there is limited research examining how they moderate the emotional expression of product versus service reviews. This study aims to fill this gap by investigating the extent to which cultural backgrounds shape the sentiment and emotional intensity of customer reviews based on purchase category.

Another unresolved question in sentiment analysis research is whether gender plays a significant role in moderating the emotional expressions found in product and service reviews. Prior studies suggest that men and women process emotions differently due to biological, psychological, and social factors, influencing how they articulate satisfaction and dissatisfaction in consumer contexts (Brody, 2008; Fischer & Manstead, 2000). Women have been found to express emotions more vividly in online communication, while men tend to adopt a more neutral tone (Purnawirawan et al., 2015; Thelwall et al., 2010). However, it remains unclear whether this gendered pattern holds across different types of online reviews, particularly in distinguishing between product and service evaluations. Given the potential implications for personalized marketing strategies, this study seeks to determine whether gender differences influence how consumers express emotions in product versus service reviews across cultural contexts.

Overall, despite the growing interest in sentiment analysis and customer review mining, there remains a lack of conceptual frameworks that integrate both category-based differences (products vs. services) and cultural variations in emotional expression. Existing studies have focused predominantly on polarity classification (positive, neutral, negative) and aspect-based sentiment detection (Liu et al., 2012; Pang & Lee, 2008), but few have explored the interactions between review content, customer expectations, and cultural influences. This study aims to bridge this gap by conceptualizing the roles of product/service categorization and cultural dimensions in shaping customer experiences and review sentiments. Specifically, this study addresses the following research questions:

1. Do customer emotional experiences differ significantly based on two categories of products and services?
2. Does the impact of category differences on customer emotional experiences vary by gender and across cultures?

The remainder of this paper is structured as follows: Section 2 provides a comprehensive literature review on category-based differences in customer emotional experiences and the moderating roles of culture and gender, with hypotheses formulated at the end of each subsection. Section 3 outlines the research design and methodology, detailing the data collection and analysis processes. Section 4 presents the study's results and findings, followed by a discussion in Section 5, which explores the significance of the findings, study limitations, and directions for future research. Finally, Section 6 concludes the paper with key takeaways and implications for theory and practice.

## 2. Literature Review

This section reviews existing literature on categorical differences in customer emotional experiences, emphasizing key insights and contrasting viewpoints. It examines how product and service categories shape emotional expression in online reviews, focusing on sentiment, valence, arousal, dominance scores, and distinctions across both basic and complex emotion categories. Furthermore, it explores cultural and gender-based variations in emotional articulation, considering both traditional assumptions and shifting digital communication patterns. By identifying gaps in prior research—such as conflicting findings, methodological constraints, and the limited use of multidimensional emotion analysis—this review establishes a foundation for developing hypotheses that will be empirically tested later on.

## 2.1. Category Differences

Understanding customer emotional experiences in online reviews is a critical area of inquiry for both academic research and business practice. Prior studies have established that consumer sentiment plays a crucial role in shaping purchasing decisions, influencing brand reputation, and predicting customer satisfaction (Berger et al., 2020). Sentiment analysis techniques have been widely applied to assess the polarity of customer reviews, typically classifying them as positive, negative, or neutral (Pang & Lee, 2008). However, recent research underscores the importance of moving beyond simple sentiment classification to examine more nuanced emotional dimensions, such as valence (pleasure), arousal (intensity), and dominance (control) (Preoţiuc-Pietro et al., 2016; Russell, 2003). While sentiment analysis has traditionally treated product and service reviews collectively, emerging evidence suggests that emotional experiences may systematically differ based on the nature of the purchase, warranting further theoretical and empirical investigation (Rasappan et al., 2024).

Theoretical perspectives from consumer psychology and behavioural economics provide a strong foundation for understanding the emotional differences between product and service reviews. One of the earliest and most influential theories, Maslow's hierarchy of Needs suggested that consumer behaviour is driven by a hierarchy of psychological and physiological expectations (McLeod, 2007). Products, particularly utilitarian goods, often fulfil basic physiological and safety needs, whereas services, particularly experience-based and relational services, tend to fulfil higher-order social and self-actualization ones (Hoffman & Novak, 2018). This distinction suggests that customer emotions associated with services may be more intense, given their connection to social and psychological fulfilment.

Another relevant theoretical framework is Zeithaml and Parasuraman (2004)'s goods-services continuum, which differentiates between goods and services based on their tangible versus intangible characteristics. This model posits that services involve greater variability, perishability, and inseparability of production and consumption, leading to heightened emotional responses compared to product purchases, which are often evaluated based on functional attributes such as durability, reliability, and price (Vargo & Lusch, 2004). Furthermore, theories of customer satisfaction suggest that negative service experiences provoke more intense dissatisfaction and stronger negative emotions than negative product experiences, primarily due to the personal and relational aspects of service consumption (Bateson & Hoffman, 2011).

Empirical studies reinforce these theoretical distinctions by demonstrating that emotional expressions in service reviews tend to be more polarized and emotionally charged. For example, Li et al. (2020) found that service reviews on platforms like TripAdvisor and Yelp exhibit a higher degree of emotional intensity compared to product reviews on Amazon. This pattern aligns with previous findings that service failures trigger stronger emotional responses due to higher consumer expectations for personalized interactions and seamless experiences (Sparks & Browning, 2011). Additionally, research suggests that customers rely more on experiential and affective language when reviewing services, whereas product reviews tend to focus on cognitive and attribute-based evaluations (Yin et al., 2014).

Beyond overall sentiment, researchers have explored specific emotional dimensions, such as valence, arousal, and dominance, in customer reviews. The circumplex model of the effect (Russell, 1980) posits that emotions can be mapped onto these three dimensions, offering a more granular understanding of customer emotional responses. Studies applying this model to online reviews indicate that service-related emotions tend to exhibit higher arousal and lower dominance levels, reflecting the heightened intensity and lower sense of control consumers feel in service encounters (Xu et al., 2023). Conversely, product reviews, particularly for utilitarian goods, are often characterized by lower arousal and higher dominance, as consumers evaluate them in a more rational and controlled manner (Yin et al., 2017).

Despite these insights, significant gaps remain in the literature. First, while many studies have examined sentiment polarity, relatively few have systematically integrated valence-arousal-dominance scores into sentiment analysis models. Additionally, findings on emotional expression in online reviews remain inconsistent across studies. For instance, while some research suggests that service reviews exhibit stronger emotional intensity (Li et al., 2020), other studies indicate that certain product categories, such as luxury goods, elicit emotions comparable to services due to their hedonic nature (Chitturi et al., 2008). Furthermore, emerging evidence suggests that factors such as reviewer personality traits, review context, and cultural background may moderate emotional expression in ways that have yet to be fully explored (Hong et al., 2016). Moreover, studies using different sentiment analysis techniques have produced conflicting results. Some research relying on lexicon-based approaches (e.g., Liu et al. (2012) suggests clear distinctions between product and service emotions, while others using deep learning-based methods report more complex and overlapping emotional structures (Wang et al., 2022).

In addition to sentiment polarity and dimensional affect scores, emotional categorization models have been employed to differentiate between 6 basic emotions (e.g., joy, anger, sadness) and up to 27 advanced emotions (e.g., gratitude, disappointment, nostalgia) (Cowen & Keltner, 2017). Research has demonstrated that service reviews contain a wider range of both basic and advanced emotions, with a greater prevalence of gratitude and disappointment due to the inherently relational nature of service experiences (Jang & Kim, 2011). By contrast, product reviews tend to emphasize emotions such as trust and surprise, which are linked to product reliability and performance expectations (Eslami et al., 2022). However, an integrated framework that combines these three dimensions—sentiment polarity, valence-arousal-dominance scores, and basic and advanced emotion categories—has not yet been systematically developed in the literature.

Building on these insights, this study proposes the following hypothesis:

H1: Customer emotional experiences differ significantly between product and service reviews, as measured by sentiment polarity, valence-arousal-dominance scores, and sentiment, basic, and advanced emotion categories.

## 2.2. Culture Impacts

Cultural differences have long been recognized as a critical determinant of consumer behaviour, influencing attitudes, preferences, and decision-making processes (Mesquita, 2022). These differences manifest in various ways, including how consumers express emotions in customer reviews, particularly in the context of product and service evaluations. Emotional expression in reviews is shaped by cultural norms that govern communication styles, emotional regulation, and social expectations (Hampden-Turner et al., 2020).

One of the most influential frameworks for analyzing cultural differences is Hofstede's's cultural dimensions theory, which identifies six key dimensions: individualism-collectivism, uncertainty avoidance, power distance, masculinity-femininity, long-term orientation, and indulgence-restraint. These dimensions have significant implications for consumer behaviour, particularly in the realm of emotional expression and sentiment formation (De Mooij, 2019). Research suggests that individuals from individualistic cultures (e.g., the U.S., Canada, and Western Europe) tend to exhibit more explicit and polarized emotional expressions, whereas those from collectivistic cultures (e.g., China, Japan, and Latin America) favour moderated and contextually nuanced emotional articulation (Wu et al., 2016). These differences extend to customer reviews, where individualistic consumers provide more direct feedback, while collectivistic consumers prioritize social harmony, often adopting an indirect and implicit communication style (Schuckert et al., 2015).

Empirical studies have provided compelling evidence of cultural differences in sentiment polarity and emotional expression in online customer reviews (Truong, 2024). Research on hospitality and e-commerce platforms has shown that customers from Western, low-context cultures tend to use stronger sentiment expressions, often producing extremely positive or negative reviews (Lim, 2016). In contrast, consumers from high-context cultures, such as China and Japan, often provide more balanced and neutral feedback, avoiding overtly negative sentiment to maintain interpersonal harmony (Au et al., 2014). These findings align with Hall's high-context vs. low-context communication theory, which suggests that the degree of contextual dependence in communication influences linguistic and emotional expression (Hall, 1976b).

Additionally, research has demonstrated that individualistic cultures are more likely to express high-arousal emotions (e.g., excitement, frustration), whereas collectivistic cultures tend to favour low-arousal emotions (e.g., calm happiness, restrained anger) (Xu et al., 2023). These patterns have significant implications for product versus service reviews, as service evaluations often involve interpersonal interactions that are more culturally sensitive than product evaluations (AlQahtani, 2021).

While the majority of studies support the notion that cultural factors affect emotional expression in customer reviews, conflicting evidence suggests that the relationship may not always be straightforward. Some studies indicate that globalization and increased digital exposure may be reducing cultural disparities in online communication (Pentina et al., 2018). For instance, research on multinational e-commerce platforms has found that consumers from traditionally collectivistic cultures exhibit review patterns similar to those from individualistic cultures, particularly among younger, more digitally connected demographics (Cortis, 2021).

Furthermore, studies have found that factors such as industry type, brand reputation, and review platform norms can sometimes override cultural influences, leading to more uniform sentiment expression across cultural groups (Hennig-Thurau et al., 2015).

Furthermore, uncertainty avoidance—a key cultural dimension—affects the way consumers evaluate and express opinions about products and services. High uncertainty-avoidance cultures (e.g., Japan, France) tend to be more risk-averse and provide detailed, information-rich reviews, while low uncertainty-avoidance cultures (e.g., the U.S., Sweden) rely more on affective, experience-based evaluations (Purnawirawan et al., 2015). This suggests that emotional differentiation between product and service reviews may be more pronounced in low uncertainty avoidance cultures, where consumers rely on emotional experiences to guide decision-making.

Moreover, while previous research has primarily focused on sentiment valence, fewer studies have examined the multidimensional nature of emotions in customer reviews, particularly regarding cultural moderating effects. Emotional dimensions such as valence (positive-negative), arousal (high-low), and dominance (control-submission) have been shown to vary across cultures (Fontaine et al., 2007), yet their role in shaping product versus service reviews remains underexplored. Additionally, studies on emotional vocabulary usage reveal that Western consumers tend to use more discrete, psychologically distinct emotional terms (e.g., joy, disappointment, nostalgia), whereas Eastern consumers prefer context-dependent, relational expressions (Lim, 2016). However, research has yet to systematically examine how these differences manifest in product versus service reviews, particularly in the context of advanced emotions as many as 27 (Cowen & Keltner, 2017).

Addressing this gap, this study integrates sentiment polarity, dimensional effect models (valence, arousal, dominance), and emotional categorization to comprehensively assess how cultural factors shape the emotional divergence between product and service reviews. Accordingly, this study hypothesizes that:

H2: The impact of product/service category differences on customer emotional experiences varies across cultures, as measured by sentiment, valence, arousal, and dominance scores, as well as sentiment, basic, and advanced emotion categories.

### 2.3. Gender impacts

The role of gender in emotional experiences has been extensively studied across disciplines, including psychology, marketing, and consumer behaviour. A growing body of research suggests that men and women differ significantly in how they perceive, express, and respond to emotions (Brody & Hall, 2010). These differences stem from a complex interplay of biological, psychological, and sociocultural factors, shaping individuals' emotional experiences across various contexts (McFarlane, 2024). Despite the extensive research on gender and emotion, limited studies have directly examined how these differences influence consumer-generated reviews, particularly in the context of product versus service evaluations.

Psychological research has long established that men and women differ in their emotional processing, with women reporting greater emotional intensity across various states (Löffler &

Greitemeyer, 2023). The Gender Socialization Theory (Eagly & Wood, 2012) posits that societal norms encourage women to express emotions more openly, particularly in interpersonal contexts, whereas men are socialized to exhibit emotional restraint. These differences manifest in emotional expression, including written and verbal communication, suggesting that consumer reviews may also reflect these gender-based patterns.

The Biopsychosocial Model of Emotion (Barrett & Lida, 2024; Barrett et al., 2007) further explains that gender differences in emotional experiences arise from an interaction of biological factors (e.g., hormonal influences), psychological traits (e.g., empathy levels), and socialization processes. Given that consumer reviews serve as a communicative tool, these gendered emotional tendencies likely shape the way men and women articulate their sentiments toward products and services.

Sentiment analysis studies have revealed gender-based variations in emotional expression in digital communication, including product reviews, social media posts, and consumer feedback (Thelwall, 2018). Women's reviews tend to exhibit more polarized sentiment—either highly positive or highly negative—compared to men, who often provide more neutral evaluations. This aligns with the Polarity Hypothesis (Tannen, 1990), which suggests that women are more likely to use emotionally expressive language in social interactions, whereas men tend to adopt a more detached and objective tone (Truong, 2016). Applying this framework to consumer reviews suggests that women may provide more intense emotional feedback, particularly for service experiences, which involve human interaction. In contrast, men may exhibit stronger emotional reactions in product categories where functional performance is critical (Zhang et al., 2023).

Given that consumer reviews represent a rich source of emotional expression, it is crucial to understand how gender moderates the way individuals react to and describe their experiences. Prior research suggests that women's feedback tends to be more emotionally expressive and nuanced, whereas men's reviews are often more neutral and focused on functional attributes (Meyers-Levy & Loken, 2015). However, conflicting findings exist, with some studies suggesting that men may exhibit stronger emotional expressions in contexts where they are personally invested (Thelwall, 2018). This study seeks to address these inconsistencies by examining how gender moderates the impact of product and service category differences on consumer emotional experiences.

Despite extensive research on gender differences in emotion, several gaps remain in the literature. First, that is the lack of integration between emotional dimensions and consumer reviews: While studies have examined gender differences in sentiment analysis, few have explored how these differences manifest across valence, arousal, and dominance in product and service evaluations. Second, there are inconsistencies in emotional expression by gender: Some studies suggest that women are more emotionally expressive in online communication (Macedo & Saxena, 2024), whereas others indicate that men can be equally expressive when the subject matter is personally relevant (Thelwall, 2018). Another gap is the limited research on advanced emotional categories: Most research has focused on basic emotions (e.g., happiness, anger), but little attention has been given to complex emotions (e.g., nostalgia, pride, guilt) in consumer reviews (Cowen & Keltner, 2017). Lastly, while gender differences in emotion have been studied in Western contexts, cross-cultural variations in emotional expression remain underexplored (Kanwal et al., 2022).

Addressing these gaps will provide a more nuanced understanding of how gender influences consumer emotional experiences and how businesses can tailor their marketing strategies accordingly. Given the theoretical and empirical insights discussed above, this study proposes the following hypothesis:

H3: The impact of product/service category differences on customer emotional experiences varies by gender, as measured by sentiment, valence, arousal, and dominance scores, as well as sentiment polarity, basic, and advanced emotion categories.

By examining these hypotheses, this study seeks to provide a more comprehensive understanding of category and culture-based emotional expression in digital communication and consumer behaviour. The next section will introduce the conceptual model underpinning this research, outlining the hypothesised direct and moderating effects, the data collection and analysis methods, and the rationale for selecting these techniques to evaluate the proposed hypotheses.

## 3. Methodology

This section will introduce the conceptual model for the study, followed by a discussion on the data collection process and the statistical techniques employed for analysis.

### 3.1. Conceptual Model

The conceptual model as shown in Figure 1 illustrates the relationships between Category, Culture, and Gender in shaping Emotional Experiences. The model outlines three hypotheses (H1, H2, and H3) that describe how these factors interact and influence different aspects of emotional responses. Emotional experiences are categorized into three components: Sentiment, Dimensional Emotions, and Categorial Emotions, which are measured through various scores and classifications.

The Category variable differentiates between classifications of offerings, specifically products and services. According to the model, Category is hypothesized to have a direct impact on Emotional Experiences (H1), implying that consumers exhibit distinct emotional responses depending on whether they engage with a product or a service. This distinction suggests that the nature of the offering itself influences the intensity, valence, and complexity of emotional reactions in consumer evaluations.

Culture is included as a moderating factor in this model (H2), influencing how different categories impact emotional experiences. Cultural background significantly affects emotional perception, expression, and categorization. Certain emotions may be more prominent or differently interpreted across cultures. For instance, collectivist cultures may emphasize emotions related to social harmony, while individualistic cultures may highlight emotions related to personal achievement.

Gender is also considered an influential factor (H3), modifying how categories affect emotional responses. Research suggests that men and women may differ in emotional sensitivity, expression, and regulation. This hypothesis assumes that gender plays a role in shaping emotional experiences, potentially leading to differences in emotional scores across different categories.

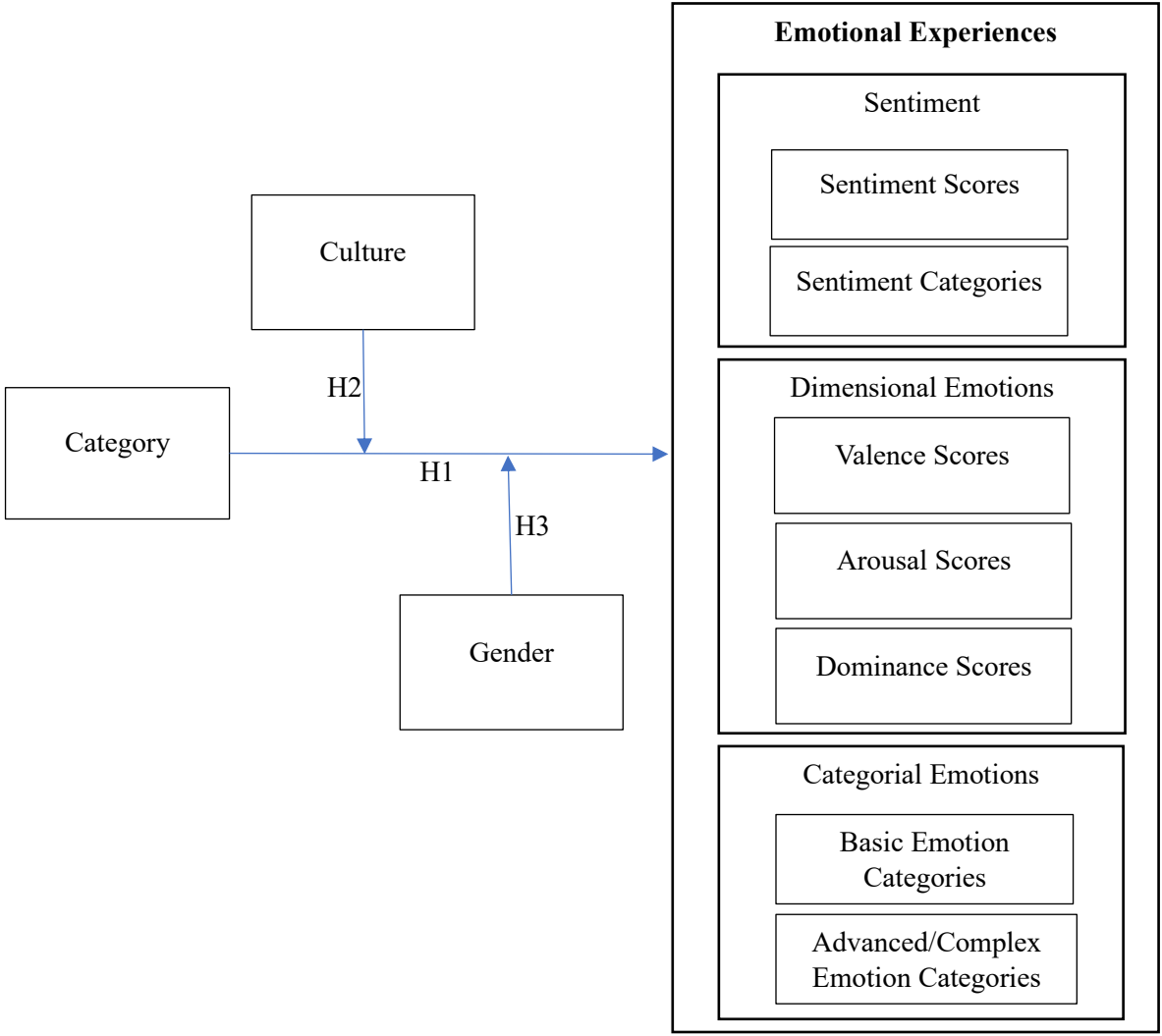

*Figure 1: The conceptual model of this study*

The Emotional Experiences section in the model is divided into three key components. The Sentiment aspect includes Sentiment Scores and Sentiment Categories, which classify emotions into positive, negative, or neutral valence. This approach is often used in sentiment analysis to determine the overall affective tone of a given stimulus.

The Dimensional Emotions framework uses Valence Scores, Arousal Scores, and Dominance Scores to represent emotions in a continuous space. Valence refers to the positivity or negativity of an emotion, arousal indicates its intensity, and dominance reflects the level of control or submissiveness associated with the emotion. This approach aligns with the Valence-Arousal-Dominance (VAD) model, which is widely used in emotion research.

The Categorial Emotions section classifies emotions into Basic Emotion Categories (such as joy, sadness, anger, fear, disgust, and surprise) and Advanced/Complex Emotion Categories, which may include emotions like nostalgia, pride, or envy. These categories help differentiate between fundamental emotional states and more nuanced affective experiences influenced by cognitive and social factors.

This conceptual model basically explores how Category, Culture, and Gender influence Emotional Experiences, outlining three hypotheses (H1, H2, and H3) that describe their interactions. Category (products vs. services) is proposed to have a direct impact on emotional responses (H1), suggesting that different offerings evoke distinct levels of emotional intensity and complexity. Culture acts as a moderator (H2), shaping how consumers from different cultural backgrounds perceive and express emotions, with collectivist cultures emphasizing social harmony and individualistic cultures focusing on personal achievement. Gender (H3) further modifies emotional responses, as research suggests men and women differ in emotional sensitivity, expression, and regulation. Emotional Experiences are classified into Sentiment, Dimensional Emotions, and Categorial Emotions, assessed through sentiment scores and classifications, which help determine the affective tone of consumer evaluations.

## 3.2. Data collection

In pursuit of the mentioned objectives, this study proposed a comprehensive method for extracting and categorizing customer emotions, consisting of several steps.

In the initial stage of gathering customer experiences, this study focused on collecting reviews from open-source platforms. A key selection criterion was the inclusion of category, cultural and gender background information, as these demographic details are essential for analyzing the effect of category and the moderating effects of culture and gender on emotional responses. By prioritizing datasets that provide this information, the study ensures a comprehensive and diverse representation of consumer perspectives across platforms, industries, and product categories.

Several publicly available datasets provide valuable insights into product and service reviews, facilitating research on consumer sentiment and emotional expression. These datasets often include user-generated reviews from e-commerce platforms, online marketplaces, and service-based websites, capturing diverse consumer experiences across different categories. Examples include the Amazon Reviews Dataset, which contains millions of product reviews with ratings and textual feedback, and the Yelp Dataset, which focuses on service-related experiences in restaurants, hotels, and other businesses. Additionally, datasets such as TripAdvisor Reviews and Google Reviews offer sentiment-rich textual data specifically for hospitality and local services (Haque et al., 2018). These resources enable researchers to analyze emotional patterns, sentiment trends, and category-based differences in consumer evaluations, providing a foundation for studies on how product and service reviews differ in emotional intensity, valence, and complexity.

Despite the abundance of online customer review datasets as shown above, a significant challenge was the lack of comprehensive demographic data. While some datasets included gender information, they often lacked cultural or nationality details, and vice versa. This limitation necessitated an extensive search and meticulous filtering process to identify datasets meeting the study's criteria. The difficulty in locating datasets that are both content-rich and demographically detailed highlights a broader gap in consumer feedback data availability.

Following an exhaustive search, six datasets were identified that satisfied the necessary conditions, as shown in Table 1. These datasets span various industries, offering a broad perspective on consumer behaviour in different contexts. The selected datasets include FashionNova (fashion), Amazon Reviews (electronics), Celsius Network (cryptocurrency trading), ASOS TrustPilot (cosmetics), Qatar Airways (airline services), and La Veranda (hospitality).

The FashionNova Customer Review Dataset contains feedback from an online fashion retailer, encompassing textual reviews, star ratings, timestamps, and demographic attributes like gender, age, and location. These reviews cover clothing, accessories, and footwear, offering insights into product quality, fit, and customer satisfaction. Similarly, the Amazon Reviews Dataset provides extensive consumer-generated reviews across various electronics products, featuring star ratings, written feedback, and metadata such as review dates and reviewer profile details (Haque et al., 2018).

The Celsius Network Customer Review Dataset includes user feedback on the cryptocurrency platform's usability, interest rates, security, and customer support. This dataset features textual reviews, ratings, and occasional demographic details (AlQahtani, 2021). Meanwhile, the ASOS TrustPilot Customer Review Dataset captures customer experiences with ASOS products and services, focusing on fit, style preferences, and delivery, along with demographic markers like age, gender, and location (Parker & Alexander, 2022) .

The Qatar Airways Customer Review Dataset compiles passenger feedback on in-flight services, comfort, punctuality, and customer support. Reviews include star ratings, text comments, and traveler demographics such as nationality, age, and travel class, making it a valuable resource for assessing sentiment and service quality in the aviation sector (Hamad MA Fetais et al., 2021). Likewise, the La Veranda Customer Review Dataset contains restaurant reviews evaluating food quality, ambiance, service, and pricing. Some records also include details about visit frequency and group size (Kondopoulos, 2014).

*Table 1: Datasets with gender and culture information*

| No | Dataset | Category |
|----|---------|----------|
| 1 | Fashionnova | Product |
| 2 | Amazon Reviews | Product |
| 3 | Celsius Network | Service |
| 4 | ASOS TrustPilot | Product |
| 5 | Qatar Airlines | Service |
| 6 | La Veranda | Service |

The inclusion of diverse datasets ensures a broad spectrum of communication styles and contextual influences, minimizing biases that could arise from overrepresentation in specific domains (Hewson et al., 2016). This diversity strengthens the study's ability to assess gender- and culture-driven variations in emotional responses across different industries.

### 3.3. Sentiment and Emotion Detection

After collecting customer reviews, advanced machine learning models was employed to analyze sentiment, valence, arousal, and dominance scores. Additionally, the reviews will undergo detailed classification to identify both basic and advanced emotions along with their intensities.

Multiple machine learning models are used to examine different aspects of emotional expression. The first step involves converting textual data into sentiment scores using tabularisai/multilingual-sentiment-analysis, a model designed to classify sentiment intensity across various languages. By assigning a numerical sentiment score to each review, this model effectively distinguishes between positive, negative, and neutral sentiments. Its multilingual capability ensures robust sentiment evaluation across diverse linguistic and cultural contexts, making it particularly useful for cross-cultural studies where traditional sentiment models may fall short (Rasappan et al., 2024).

For sentiment classification, VADER is utilized to categorize text into sentiment groups (positive, negative, and neutral). Unlike conventional sentiment analysis tools, VADER considers contextual intensities, punctuation, capitalization, and even emoticons, making it highly effective for analyzing informal and opinion-rich texts, such as customer reviews (Hutto & Gilbert, 2014).

To measure valence (emotional tone), arousal (intensity), and dominance (control level) in consumer feedback, the study employs hplisiecki/word2affect_english. This model assigns numerical values to these three emotional dimensions, providing a more nuanced understanding of consumer emotions than traditional sentiment analysis (Plisiecki & Sobieszek, 2024). By evaluating emotions on a 3D scale, the study can capture subtle differences in emotional expression across gender and cultural groups.

For identifying basic emotions such as happiness, sadness, anger, fear, surprise, and disgust, the study uses bhadresh-savani/bert-base-uncased-emotion. This transformer-based model is trained on large datasets to ensure high accuracy in emotion classification (Khalili et al., 2022). Analyzing basic emotions allows for comparisons between different consumer groups, shedding light on how gender and cultural backgrounds influence emotional expression.

To capture complex emotions, the study utilizes SamLowe/roberta-base-go_emotions, a model fine-tuned on the GoEmotions dataset (Demszky et al., 2020). This dataset includes 27 distinct emotion labels, such as admiration, amusement, pride, disappointment, and gratitude. By incorporating advanced emotion classification, the study provides a more granular view of consumer sentiment, uncovering patterns that basic models might overlook. This level of analysis is particularly valuable for understanding nuanced emotional expressions across demographics, an area where previous research has been limited.

By leveraging these models, the study ensures a comprehensive, multidimensional analysis of consumer emotions, offering deeper insights into how gender and cultural factors shape emotional responses in digital communication.

### 3.4. Statistical techniques

After the machine learning model identified and categorized emotions, statistical techniques were applied to analyze how emotional expressions varied across cultural and gender groups. Descriptive statistics, including means and standard deviations, were first used to assess data distribution and ensure normality. The dataset, consisting of a relatively equal representation of product and service reviews, male and female consumers, as well as Western and Eastern reviews, enabled balanced comparisons. Additionally, key emotional metrics—sentiment scores, as well as valence, arousal, and dominance scores—were evaluated for skewness and kurtosis, confirming that their distributions fell within acceptable statistical limits.

The reliability of the analysis was strengthened by the use of a large dataset comprising many customer reviews, which helped minimize biases and ensure a more representative sample. Normality was further supported by the even distribution of sentiment categories and both basic and complex emotion classifications. These preliminary statistical checks established a solid foundation for subsequent inferential analyses, allowing for more precise comparisons of emotional expression across demographic and cultural segments.

To further investigate differences in emotional expression across cultural and gender groups, inferential statistical techniques were employed. Univariate tests were conducted to determine whether significant differences existed in the distribution of emotions between these groups. Additionally, multivariate analysis techniques provided deeper insights into the complex relationships between culture, gender, and emotional intensity. Specifically, Multivariate Analysis of Variance (MANOVA) was used to simultaneously examine multiple dependent variables, such as sentiment scores and emotion categories, across cultural and gender groups. This comprehensive approach allowed for a more holistic assessment of how these factors interact (Saunders, 2015).

Moreover, a moderation regression test using Hayes' PROCESS macro was performed to investigate whether culture moderated the effect of gender on emotional expression (Hayes, 2017). This analysis helped determine whether cultural differences amplified, diminished, or altered gender-based emotional patterns, offering a more nuanced perspective on how emotions are shaped by both demographic factors.

The following section presents the key statistical findings and their implications, providing deeper insights into the interplay between culture, gender, and emotional expression.

## 4. Results and Findings

This section presents the findings from the multivariate and moderation regression analyses, organized based on the results of the hypothesis tests. Additionally, any findings beyond the scope of the hypotheses will be discussed further.

## 4.1. Hypothesis 1 Testing

The hypothesis testing results presented in Table 2 reveal significant findings regarding the relationship between emotional categories and key emotional metrics, such as sentiment score, valence, arousal, and dominance. The analysis demonstrates that category type has a statistically significant effect on sentiment scores, with a Type III Sum of Squares value of 70.343 and an F-value of 259.398 ($p < .001$). This indicates that different emotional categories exert a meaningful influence on the overall sentiment expressed in customer reviews. The significance of this result suggests that emotional category distinctions play a crucial role in shaping sentiment intensity, reinforcing the importance of categorizing emotions accurately in consumer analysis.

However, when analyzing the relationship between emotional category and sentiment category classification, the results show no statistical significance ($p = .614$). With a low F-value of 0.254, this finding suggests that emotional categories do not meaningfully impact sentiment classification into positive, negative, or neutral labels. This result implies that while emotional categories influence sentiment intensity (as indicated by the sentiment score analysis), they may not necessarily determine whether an emotion falls into broad sentiment classifications. The lack of significance in this case highlights potential limitations in sentiment classification models, which may not capture the nuanced effects of emotion categorization on consumer sentiment.

Regarding the valence score, a statistically significant effect of category was observed, with a Type III Sum of Squares value of 0.228 and an F-value of 7.234 ($p = .007$). This result indicates that emotional category distinctions contribute to variations in valence, which represents the positivity or negativity of an emotion. Although the effect size is relatively small, the statistical significance suggests that categorizing emotions can provide insights into the valence of consumer sentiment, supporting the use of valence as a key emotional metric in sentiment analysis.

*Table 2: Hypothesis 1 testing result*

| Independent Variable | Dependent Variable | Type III Sum of Squares | df | Mean Square | F | Sig. |
|---|---|---|---|---|---|---|
| Category | Sentiment Score | 70.343 | 1 | 70.343 | 259.398 | .000 (***) |
| | Sentiment Category | .466 | 1 | .466 | .254 | .614 |
| | Valence Score | .228 | 1 | .228 | 7.234 | .007 (***) |
| | Arousal Score | 1.168 | 1 | 1.168 | 291.575 | .000 (***) |
| | Dominance Score | .123 | 1 | .123 | 11.965 | .001 (***) |
| | Basic Emotion Category | .265 | 1 | .265 | .245 | .621 |
| | Complex Emotion Category | 2333.084 | 1 | 2333.084 | 36.453 | .000 (***) |

Arousal, which measures emotional intensity, exhibited a strong relationship with emotional category, as evidenced by an F-value of 291.575 ($p < .001$). The Type III Sum of Squares value of 1.168 suggests that emotional categories significantly impact the arousal levels expressed in customer reviews. This finding aligns with previous research indicating that emotions vary in intensity depending on their classification, emphasizing the importance of measuring arousal alongside sentiment and valence for a more comprehensive understanding of emotional expression in consumer feedback.

Similarly, dominance scores were significantly affected by emotional category, with a Type III Sum of Squares value of 0.123 and an F-value of 11.965 ($p = .001$). This result indicates that emotional category distinctions contribute to variations in the perceived level of control or dominance within consumer sentiment. The significance of this relationship suggests that certain emotions, depending on their classification, may evoke stronger or weaker perceptions of control, which can influence how consumers express their experiences and interactions with a brand or product.

Interestingly, when analyzing the relationship between emotional category and basic emotion classification, no significant effect was observed ($p = .621$). With an F-value of 0.245, this suggests that emotional categories do not strongly impact how emotions are classified into basic emotions such as happiness, sadness, or anger. This may indicate that basic emotions are relatively stable across categories and that more granular emotional distinctions are necessary to capture meaningful differences in emotional expression.

In contrast, the relationship between emotional category and complex emotion classification was highly significant, with a Type III Sum of Squares value of 2333.084 and an F-value of 36.453 ($p < .001$). This strong effect suggests that emotional categories are crucial in distinguishing between complex emotions, such as admiration, gratitude, or disappointment. This finding underscores the importance of using advanced emotion classification models to capture the depth and complexity of consumer emotions, which may not be fully represented by basic sentiment or emotion classifications alone.

The differences between product and service categories in customer emotional experiences are particularly pronounced when analyzing complex emotion categories, as illustrated in Figure 2. The distribution of emotions between the two categories reveals that services tend to elicit a higher proportion of complex emotions compared to products. This is evident in emotions such as love, excitement, amusement, optimism, and admiration, where service-related experiences dominate. The nature of services, which often involves human interaction, personalized experiences, and direct engagement, likely contributes to the heightened emotional responses observed in these categories.

Conversely, product-related experiences generate a greater share of emotions like relief, pride, and caring. These emotions suggest that customers feel reassured and satisfied when a product meets their expectations, provides a solution to a problem, or enhances their daily lives. Unlike services, which often require real-time performance and interpersonal exchanges, products offer a tangible and often long-term source of utility. The prevalence of relief in product experiences indicates that customers may associate products with problem-solving and functional reliability, whereas services tend to evoke more dynamic emotional responses.

Negative emotions also exhibit interesting distinctions between the two categories. Disgust, disappointment, disapproval, and sadness are more strongly associated with services than products. This suggests that negative service experiences, such as poor customer service, unmet expectations, or service failures, lead to stronger emotional reactions than product-related dissatisfaction. Unlike products, where dissatisfaction might be mitigated through returns or replacements, service failures are often more immediate and harder to rectify, leading to stronger negative sentiment.

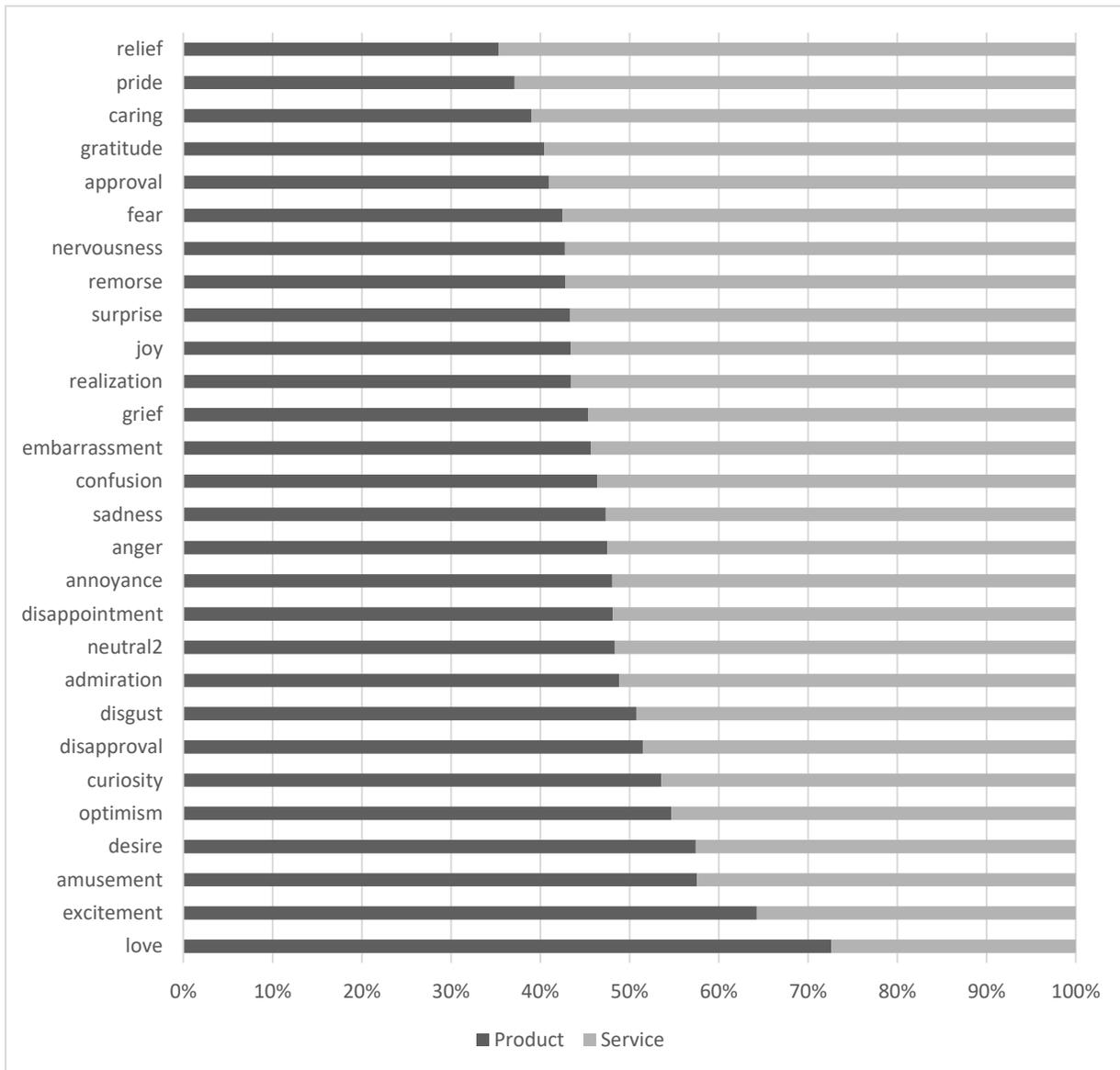

*Figure 2: Categorical differences in advanced emotion categories*

Furthermore, emotions related to uncertainty, such as confusion and embarrassment, are also more prevalent in service experiences. This trend suggests that customers may feel more vulnerable or unsure when navigating service interactions, especially in cases where processes are complex or expectations are unclear. In contrast, products generally provide a more straightforward experience, reducing the likelihood of these emotions. The higher presence of fear and nervousness in service-related experiences further underscores the emotional sensitivity surrounding service engagements, where trust, competence, and reliability play critical roles.

Overall, the distinctions in emotional experiences between products and services highlight the fundamental differences in how consumers interact with and perceive these two categories. While products tend to elicit emotions related to reliability, functionality, and relief, services generate a broader range of emotional responses, particularly in complex and interpersonal emotions. This suggests that businesses in the service industry must be especially mindful of emotional management, ensuring that customer interactions foster positive experiences while mitigating the risks of strong negative emotions.

## 4.2. Hypotheses 2 & 3 Testing

Hypothesis 2 and Hypothesis 3 aimed to examine the moderating effect of Gender and Culture variables and identify the moderating factor in this relationship. Graphs will be plotted to visually illustrate the moderating effect of one factor on another.

### 4..3.1. Multivariate Analysis

The hypothesis testing results presented in Table 3 reveal significant interactions between category and culture on various sentiment-related measures, while interactions between category and gender generally show non-significant effects. The interaction between category and culture on sentiment score exhibits a highly significant result ($F = 89.660$, $p < .001$), indicating that cultural differences play a substantial role in shaping sentiment responses across categories. This suggests that sentiment scores are significantly influenced by the interaction between cultural background and the category of interest, emphasizing the importance of cultural considerations when analyzing sentiment in different contexts.

Similarly, the interaction between category and culture on sentiment category is also highly significant ($F = 12.686$, $p < .001$). This result suggests that different cultural groups categorize sentiments differently depending on the category they are evaluating. It highlights potential variations in emotional expression and interpretation across cultures, reinforcing the need for culturally tailored sentiment analysis models. Additionally, the significant interaction effect on valence scores ($F = 49.698$, $p < .001$) further supports the notion that cultural factors influence how positive or negative a sentiment is perceived in different categories.

The analysis of arousal scores, however, reveals a non-significant interaction between category and culture ($F = 0.224$, $p = .799$). This suggests that cultural differences do not significantly impact the arousal dimension of sentiment within different categories. Unlike valence, which is strongly tied to cultural perception, arousal levels might be more universally consistent across cultures when evaluating different categories. On the other hand, the interaction between category and culture on dominance scores is highly significant (F = 22.877, p < .001), suggesting that cultural background significantly influences the degree of control or dominance experienced in response to different categories.

*Table 3: Gender and Cultural moderating effects*

| Independent Variable | Dependent Variable | Type III Sum of Squares | df | Mean Square | F | Sig. |
|---|---|---|---|---|---|---|
| Category * Culture | Sentiment Score | 48.628 | 2 | 24.314 | 89.660 | .000 (***) |
|  | Sentiment Category | 46.491 | 2 | 23.245 | 12.686 | .000 (***) |
|  | Valence Scores | 3.131 | 2 | 1.565 | 49.698 | .000 (***) |
|  | Arousal Score | .002 | 2 | .001 | .224 | .799 |
|  | Dominance Score | .470 | 2 | .235 | 22.877 | .000 (***) |
|  | Basic Emotion Category | 6.896 | 2 | 3.448 | 3.188 | .041 (***) |
|  | Advanced Emotion Categories | 116.913 | 2 | 58.456 | .913 | .401 |
| Category * Gender | Sentiment Score | .016 | 1 | .016 | .058 | .809 |
|  | Sentiment Category | 1.570 | 1 | 1.570 | .857 | .355 |
|  | Valence Scores | 1.641E-5 | 1 | 1.641E-5 | .001 | .982 |
|  | Arousal Score | 5.689E-5 | 1 | 5.689E-5 | .014 | .905 |
|  | Dominance Score | .030 | 1 | .030 | 2.890 | .089 |
|  | Basic Emotion Category | 1.536 | 1 | 1.536 | 1.420 | .233 |
|  | Advanced Emotion Categories | .001 | 1 | .001 | .000 | .997 |

The results also show a significant interaction between category and culture on basic emotion categories (F = 3.188, p = .041), indicating that cultural differences influence the way individuals categorize basic emotions within different categories. However, the interaction effect on advanced emotion categories is not significant (F = 0.913, p = .401), suggesting that while basic emotions may vary in expression across cultures, more complex emotional classifications remain relatively stable regardless of cultural background. This distinction highlights the complexity of cultural influence on emotional perception, with basic emotions being more susceptible to cultural variation than advanced emotions.

In contrast to the significant effects observed for category and culture, the interaction between category and gender on sentiment-related measures yields consistently non-significant results. The interaction effect on sentiment score is negligible (F = 0.058, p = .809), indicating that gender does not significantly moderate sentiment differences across categories. Similarly, sentiment category (F = 0.857, p = .355), valence scores (F = 0.001, p = .982), and arousal scores (F = 0.014, p = .905) all exhibit non-significant effects, further reinforcing the notion that gender differences do not substantially influence sentiment responses across categories.

Although the interaction effect of category and gender on dominance scores is marginally close to significance (F = 2.890, p = .089), it remains non-significant, suggesting that any observed differences in perceived dominance between categories and genders are not strong enough to be statistically meaningful. Similarly, both basic emotion category (F = 1.420, p = .233) and advanced emotion category (F = 0.000, p = .997) show non-significant interactions, indicating that gender does not significantly influence how emotions are categorized within different categories. Overall, the findings suggest that while cultural differences play a significant role in shaping sentiment and emotional responses across categories, gender differences do not exhibit the same level of influence.

### 4.3.2. Moderation Regression Analysis

The moderation regression analysis utilizes Hayes' PROCESS macro to continue assessing the moderating effects of Gender and Culture on their relationship with Customer Emotional Experiences. Visual representations will be used to illustrate these moderating effects more clearly.

*Moderating the effect on Sentiment Score*

Figure 3 illustrates the interaction between category type (Product vs. Service) and cultural background (Eastern vs. Western) on sentiment scores. The straight line represents the Eastern cultural group, while the dash line represents the Western cultural group. A notable pattern emerges: while the Western group maintains a relatively stable sentiment score across both product and service categories, the Eastern group shows a significant increase in sentiment from products to services. This indicates that cultural background moderates the relationship between category type and sentiment, leading to varying emotional responses based on cultural context.

For the Eastern cultural group, sentiment toward products starts at a lower level but rises substantially when shifting to services. This suggests that individuals from Eastern cultures may associate more positive emotions with services than with products. In contrast, the Western cultural group exhibits a relatively flat trend, implying that their sentiment remains consistent regardless of whether they are evaluating a product or a service. This stability suggests that Western consumers may not differentiate as strongly between the two categories in terms of sentiment.

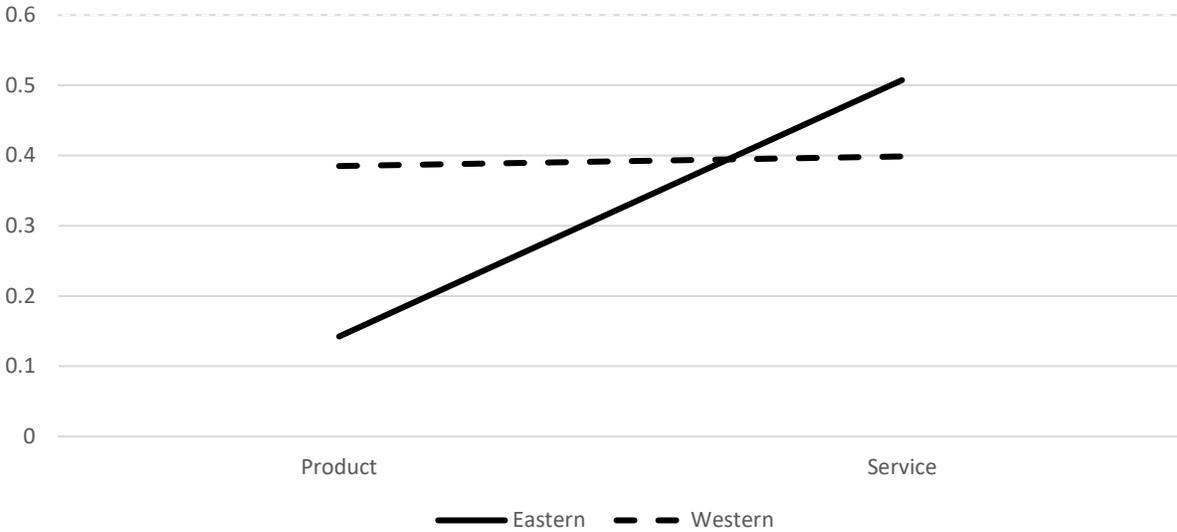

*Figure 3: Moderating effects on the sentiment score*

The interaction effect observed in the graph has important implications for marketing strategies. Since Eastern consumers exhibit significantly higher sentiment toward services, businesses targeting this demographic should emphasize service-related aspects in their campaigns. This could include highlighting customer service quality, personalized experiences, or after-sales support. On the other hand, since Western consumers do not show a strong variation in sentiment between products and services, marketing strategies for this group may not need to prioritize one category over the other but rather focus on other influential factors such as quality, brand perception, or value proposition.

*Moderating the effect on Valence score*

Figure 4, similarly, presents an interaction effect between category type (Product vs. Service) and cultural background (Eastern vs. Western) on sentiment scores. The straight line represents the Eastern cultural group, while the dash line represents the Western cultural group. The key trend in the graph shows that sentiment scores increase for the Eastern group when shifting from products to services, whereas sentiment scores decrease for the Western group. This indicates that cultural background significantly influences how consumers perceive products versus services.

For the Eastern cultural group, sentiment toward products starts at a relatively lower level but increases when transitioning to services. This suggests that individuals from Eastern cultures may have a more positive emotional response toward services compared to products. In contrast, the Western cultural group begins with a higher sentiment score for products, but their sentiment score declines when evaluating services. This pattern implies that Western consumers may place greater emotional value on products than on services.

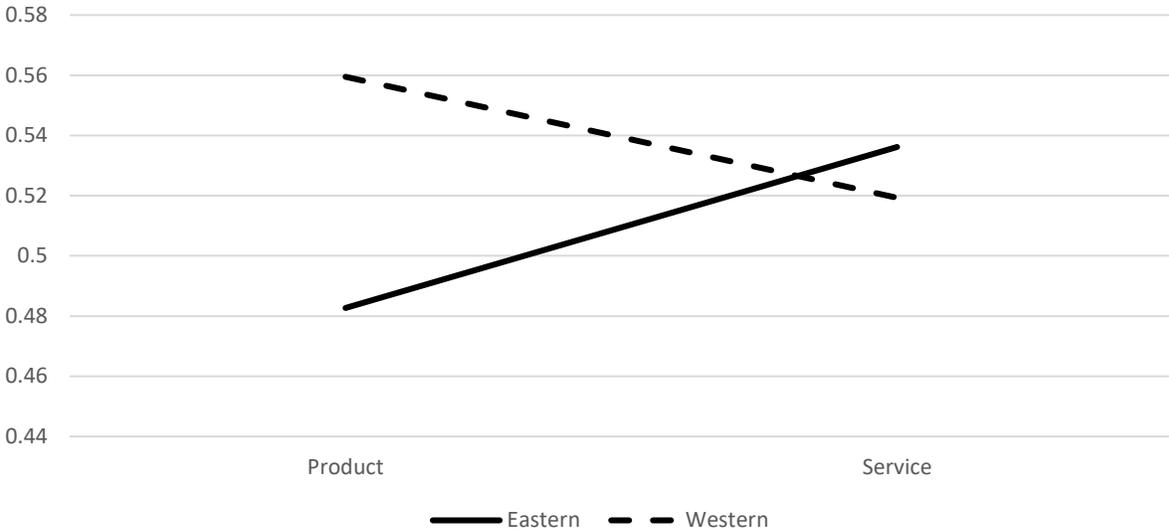

*Figure 4: Moderating effects on the valence score*

The interaction effect revealed in the graph has strategic implications for businesses and marketers. Since Eastern consumers exhibit stronger sentiment toward services, companies targeting this demographic should emphasize service-related features, such as quality customer service, personalization, and post-purchase support. Conversely, since Western consumers show higher sentiment toward products, marketing efforts for this group should focus on product-related benefits, such as superior quality, innovation, and tangible value. Understanding these cultural differences can help businesses craft more effective and culturally tailored advertising strategies.

*Moderating the effect on Dominance Score*

Figure 5 visually illustrates the interaction effect between category type (Product vs. Service) and cultural background (Eastern vs. Western) on sentiment scores. The straight line represents the sentiment trend for the Eastern cultural group, while the dash line represents the Western cultural group. The graph shows that sentiment scores increase for the Eastern group when moving from products to services, whereas the sentiment scores slightly decline for the Western group. This suggests that culture plays a significant role in shaping consumer sentiment toward different categories.

For the Eastern cultural group, sentiment toward products starts at a lower level and increases when transitioning to services. This indicates that individuals from Eastern cultures may have a more positive emotional response to services compared to products. On the other hand, the Western cultural group begins with a higher sentiment score for products, but their sentiment slightly decreases for services. This trend implies that Western consumers may value products more than services, whereas Eastern consumers place greater sentiment on services.

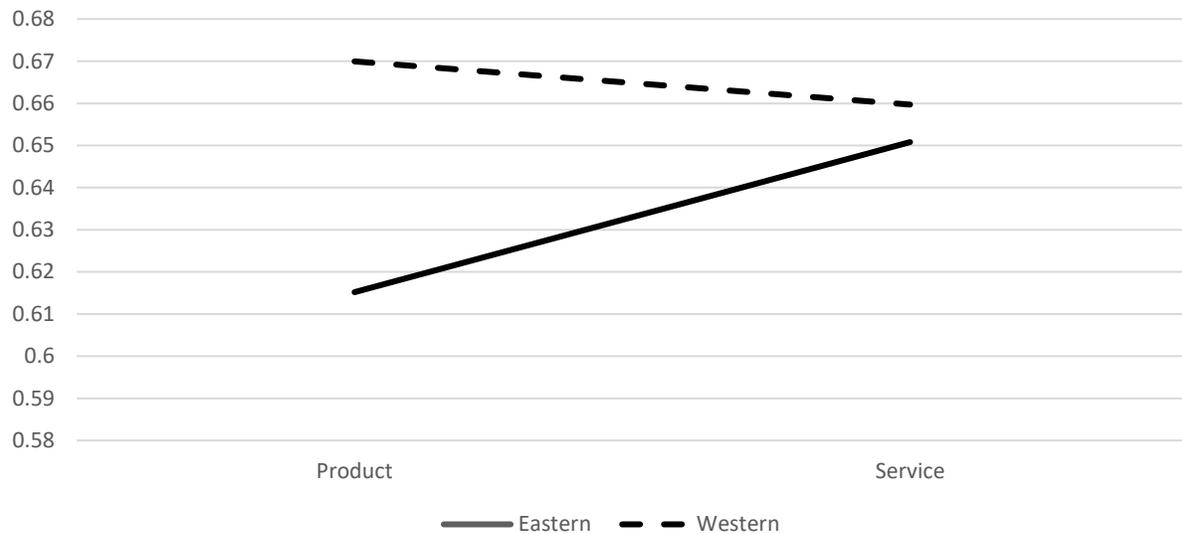

*Figure 5: Moderating effect on the dominance score*

This interaction has important implications for businesses and marketers. Since Eastern consumers demonstrate stronger sentiment toward services, companies targeting this demographic should focus on highlighting service quality, customer experience, and relationship-building aspects. Conversely, Western consumers show a preference for products, suggesting that marketing strategies for this group should emphasize product benefits such as innovation, functionality, and reliability. Recognizing these cultural differences can help businesses develop tailored marketing campaigns that resonate with their target audience.

### 4.3. Summary

Table 4 presents the hypothesis testing results for the impact of category and the interaction of category with culture on various sentiment-related dependent variables. The first hypothesis examines whether the category (Product vs. Service) significantly influences sentiment scores, sentiment categories, valence scores, arousal scores, dominance scores, and basic emotion categories. The results indicate that the category variable has a significant effect on most sentiment-related variables, suggesting that services tend to generate stronger sentiment responses compared to products. This finding aligns with the remark that sentiment scores are generally higher for services than for products, reinforcing the idea that consumers may have stronger emotional reactions to services.

The second hypothesis tests the interaction between category and culture (Eastern vs. Western) to determine whether cultural background further amplifies the differences in sentiment responses between products and services. The results indicate that the interaction effect is significant for sentiment score, sentiment category, and valence scores. This suggests that cultural background influences how consumers emotionally respond to products and services. Specifically, the remark states that the difference between sentiment responses to services and products is even more pronounced among Eastern consumers compared to Western consumers. This implies that cultural factors play a moderating role in shaping sentiment-related responses.

Examining the results further, valence scores are significantly affected by both category and the interaction between category and culture. Since valence represents the positive or negative nature of an emotion, this suggests that the emotional tone associated with services is not only higher overall but also varies significantly depending on cultural background. The finding that the effect is stronger in Eastern cultures could mean that Eastern consumers perceive a greater emotional contrast between products and services, whereas Western consumers may have a more neutral response to both categories.

Interestingly, while the category variable significantly affects the arousal and dominance scores, the interaction between category and culture does not seem to impact these variables. This suggests that while services may generally evoke higher arousal and dominance levels compared to products, these emotional dimensions are not strongly influenced by cultural differences. In other words, while culture influences the valence (positivity or negativity) of emotions toward products and services, it does not significantly alter the intensity (arousal) or sense of control (dominance) of those emotions.

Unlike culture, which significantly influences how category impacts emotional experiences, gender does not exhibit a moderating effect in this relationship. This suggests that men and women express their emotions similarly when reviewing products and services, showing no substantial differences in sentiment, emotional intensity, or complexity across categories.

*Table 4: Hypothesis testing results*

| Hypothesis | Independent variable | Sentiment Score | Sentiment Category | Valence Scores | Arousal Score | Dominance Score | Basic Emotion CCategory | Advanced Emotion Category | Remark |
|---|---|---|---|---|---|---|---|---|---|
| 1 | Category | Yes | | Yes | Yes | Yes | | Yes | Service is higher than Product |
| 2 | Category x Culture | Yes | | Yes | | Yes | | | Eastern is even more different than Western |
| 3 | Category x Gender | | | | | | | | |

Overall, these findings highlight the importance of considering both category and cultural background when analyzing consumer sentiment. Services tend to generate higher sentiment scores than products, and this effect is even stronger for Eastern consumers. These insights are particularly valuable for marketers and businesses seeking to tailor their strategies for different cultural audiences. Understanding that sentiment responses differ across cultures and product categories can help in designing more effective advertising campaigns, improving customer experiences, and enhancing brand positioning in global markets.

Although gender appears to have little effect on emotional experiences in product and service reviews, this finding still holds important implications for businesses. It suggests that companies can adopt more generalized emotional analysis strategies without needing to tailor feedback interpretation based on gender. Instead, businesses can focus on other influential factors, such as cultural differences or category-specific emotional responses, to refine their marketing and customer engagement strategies.

Based on the findings summarised in Table 4, the next section will discuss the theoretical contributions, practical implications, and limitations of the results, providing insights for future research.

# 5. Discussion
## 5.1. Theoretical contributions

The findings above align with Maslow's hierarchy of needs (McLeod, 2007), particularly in understanding how different categories of consumption—products versus services—satisfy various human needs. Maslow's framework suggests that human motivations are structured in a pyramid, starting with basic physiological needs and progressing to higher-order psychological and self-fulfillment needs. Products often fulfill lower-level needs, such as physiological (food, clothing) and safety needs (insurance, home security), whereas services are more likely to engage higher-level needs, such as social belonging, esteem, and self-actualization. The stronger emotional responses to services observed in this study support this theory, as services often involve interpersonal interactions, recognition, and self-improvement—factors that are central to Maslow's higher-tier needs (Maslow, 1943).

Furthermore, the interaction between culture and sentiment differences suggests that cultural variations influence how consumers prioritize and experience Maslow's needs. This aligns with Hofstede's cultural dimensions, particularly the individualism-collectivism and uncertainty avoidance dimensions (Hofstede, 2011). In collectivist cultures, where social harmony and group cohesion are emphasized, services may be perceived as more integral to fulfilling relational and esteem-related needs. This may explain why Eastern consumers in the study exhibited a more pronounced sentiment gap between products and services compared to Western consumers. Conversely, in individualistic cultures, where personal achievement and independence are prioritized, products that signal status or self-actualization may hold greater emotional weight than in collectivist settings. Additionally, cultures with high uncertainty avoidance may show stronger emotional responses to services that provide guidance, structure, or expertise, as these cultures tend to value predictability and trust in institutions. These observations highlight the necessity of incorporating cultural dimensions when applying Maslow's hierarchy to consumer behavior.

The study's findings also suggest that emotional responses to services may be linked to self-actualization, the highest level of Maslow's hierarchy. Services related to education, personal growth, wellness, and entertainment often help consumers achieve personal development and fulfillment, which are key components of self-actualization. The elevated sentiment scores for services align with the idea that consumers seek transformative experiences beyond material possessions. This reinforces the theoretical perspective that modern consumers, particularly in service-dominated economies, are increasingly drawn to experiences that enrich their personal and social lives rather than just acquiring products.

Moreover, the dimensions of valence, arousal, and dominance examined in this study can be mapped onto Maslow's hierarchy by considering how different emotional states correspond to various levels of need fulfilment. High valence (positive emotion) and high arousal (intensity of feeling) can be associated with esteem and self-actualization needs, where individuals feel excitement, joy, or accomplishment from engaging with services (McLeod, 2007). Meanwhile, dominance (sense of control) may relate to safety and security needs, suggesting that products providing security and stability might evoke stronger dominance-related emotional responses. These insights bridge sentiment analysis with human motivation theories, emphasizing the complex interplay between emotions and consumer needs.

By linking the study's findings to Maslow's hierarchy of needs and Hofstede's cultural dimensions, this research extends existing theories on emotional engagement in consumer behavior. It supports the idea that services, which tend to fulfill higher-order psychological and self-fulfillment needs, elicit stronger emotional responses than products, which often address fundamental or security needs. Additionally, cultural differences in sentiment responses further highlight how Maslow's hierarchy is not universal but rather shaped by sociocultural contexts, particularly the distinctions between collectivist and individualist cultures. This theoretical integration provides a more nuanced understanding of consumer sentiment and offers valuable implications for marketers aiming to align their strategies with deeper human motivations while considering cultural variations in emotional expression and decision-making.

## 5.2. Practical implications

The practical implications of this study are significant for businesses, marketers, and service providers, as the findings highlight the varying emotional responses to products and services. Since services evoke stronger emotions, businesses operating in the service industry should leverage emotional engagement strategies in their marketing and customer interactions. This could involve emphasizing personal connections, storytelling, and customer experiences that align with consumers' psychological and self-fulfilment needs. For example, brands offering educational, wellness, or entertainment services should craft messages that highlight personal growth, satisfaction, and transformative experiences rather than focusing purely on functional benefits.

Additionally, the study's findings suggest that cultural differences influence sentiment toward products and services. This has practical implications for global marketing strategies, as businesses must tailor their messaging based on cultural contexts. In Eastern cultures, where sentiment differences between products and services are more pronounced, companies should emphasize relational and experiential aspects of services, reinforcing their value in social and communal settings. Meanwhile, in Western cultures, where sentiment scores are more balanced, product marketing could incorporate elements of self-expression and achievement, resonating with individualistic values. These insights can help businesses create culturally relevant advertising campaigns that align with consumer emotions.

The results also provide valuable insights for customer experience management, particularly in service industries where emotional engagement is a key driver of customer satisfaction and loyalty. Businesses should focus on creating positive service experiences by ensuring seamless interactions, personalized customer service, and emotional touchpoints that foster a sense of belonging and appreciation (Truong, 2022). High emotional engagement can translate into customer retention, brand advocacy, and word-of-mouth referrals, making emotional sentiment analysis a crucial tool for service-oriented companies. By monitoring sentiment scores, businesses can assess the impact of customer interactions and refine their strategies accordingly.

For product-based businesses, the study suggests that while products may not elicit as strong emotional responses as services, companies can enhance engagement by integrating service elements into their offerings. For example, brands can offer post-purchase support, customer education, or loyalty programs that create an ongoing relationship with consumers. Subscription-based models, experiential marketing, and personalized product recommendations can help bridge the emotional gap between products and services, fostering deeper customer connections.

The findings also have implications for pricing strategies. Since services generate stronger emotional responses, businesses in the service industry may justify premium pricing by emphasizing the unique experiences and psychological benefits their services provide. Consumers are often willing to pay more for services that offer emotional fulfilment, such as travel, wellness, or entertainment experiences. On the other hand, product-based businesses may need to compete more on value, quality, and differentiation to enhance perceived worth, especially in competitive markets where emotional appeal is less pronounced.

Furthermore, digital marketing strategies can be refined using sentiment analysis insights from this study. Businesses can utilize AI-driven sentiment tracking tools to assess consumer reactions to advertisements, social media campaigns, and online reviews, as these tools are very popular today (Truong & Hoang, 2022). By understanding how consumers emotionally engage with their brand, companies can optimize content strategies to maximize positive sentiment and mitigate negative emotional responses. This approach is particularly useful in industries where brand perception and emotional connection significantly influence purchasing decisions.

## 5.3. Limitations and Directions for Future Research

While this study provides valuable insights into the emotional responses associated with products and services, it has certain limitations that should be addressed in future research. One key limitation is the reliance on sentiment analysis, which, although effective, may not fully capture the complexity of human emotions. Sentiment scores, valence, arousal, and dominance measures provide quantitative insights, but they may not account for deeper psychological, cultural, or contextual factors influencing emotional responses. Future research could incorporate qualitative methods, such as interviews or focus groups, to gain a more nuanced understanding of consumer sentiment and its underlying drivers.

Another limitation is the scope of cultural comparisons in this study. While the findings indicate that Eastern and Western cultures exhibit different sentiment patterns toward products and services, cultural influences are multifaceted and cannot be fully captured by broad categorizations. Future research could explore more diverse cultural contexts, including regional and subcultural differences, to refine the understanding of how cultural values, social norms, and personal experiences shape consumer sentiment. Additionally, investigating generational differences within cultures may provide further insights into evolving consumer preferences and emotional responses across age groups.

The study also primarily focuses on sentiment related to products and services but does not extensively examine other moderating factors such as brand reputation, industry type, or consumer personality traits. Future research could explore how these factors interact with sentiment to influence purchasing decisions and brand loyalty. For instance, brand trust and familiarity may mitigate or amplify emotional responses, and personality traits such as openness or neuroticism could impact how consumers perceive and react to different offerings. Expanding the research scope to include these variables could enhance the applicability of sentiment-based insights in marketing and consumer behavior studies.

Lastly, the study's findings could be further validated by employing longitudinal research designs to assess how sentiment evolves over time. Emotional responses to products and services may shift due to changing consumer expectations, technological advancements, or socio-economic trends. By conducting longitudinal studies, researchers could identify patterns in emotional engagement and sentiment shifts, allowing businesses to anticipate and adapt to evolving consumer behavior. Additionally, experimental research could help establish causal relationships between sentiment, cultural influences, and purchasing behavior, further strengthening the theoretical and practical implications of the study.

# 6. Conclusions

This study explores the relationship between category (products vs. services), culture, and gender in shaping emotional experiences in digital consumer reviews. It examines how different types of offerings evoke distinct emotional responses and whether cultural and gender differences moderate these effects.

The findings of this study provide significant insights into how consumers emotionally respond to products and services, particularly across different cultural contexts. By examining sentiment scores, valence, arousal, and dominance measures, this study highlights the variations in emotional engagement between Eastern and Western consumers. The results indicate that services tend to elicit stronger emotional responses than products, and cultural differences further shape these sentiments. These insights contribute to a deeper understanding of consumer perception, reinforcing the importance of tailoring marketing strategies to align with emotional and cultural nuances.

The study also underscores the role of emotional factors in influencing consumer behavior, linking them to established psychological frameworks such as Maslow's hierarchy of needs. As emotions play a crucial role in decision-making, businesses must consider how their offerings fulfill not only functional but also psychological and emotional needs. The distinction between products and services in eliciting different emotional responses suggests that marketers should develop targeted communication strategies that align with the emotional expectations of their audience, ensuring a more engaging and persuasive consumer experience.

From a cultural perspective, the findings emphasize that emotional responses to products and services are not universal but rather shaped by cultural values, traditions, and social norms. Eastern consumers exhibited more pronounced differences in sentiment when comparing products and services, suggesting that their emotional engagement is more context-dependent. This highlights the necessity for culturally adaptive marketing approaches, where businesses consider local preferences and emotional tendencies when designing advertisements, customer experiences, and branding strategies.

The practical implications of this study extend to multiple industries, including e-commerce, digital advertising, and brand management. By understanding the emotional impact of their offerings, companies can optimize product positioning, enhance customer engagement, and refine user experience strategies. Additionally, businesses operating in multicultural markets can use these findings to develop culturally sensitive campaigns that resonate with diverse consumer bases, ultimately improving customer satisfaction and brand loyalty.

In conclusion, this study contributes to the growing body of research on consumer sentiment by highlighting the difference between product and service reviews and the moderating effect of the culture factor. It reinforces the need for businesses to adopt emotion-driven and culturally adaptive strategies to maximize consumer engagement. By building upon these insights, future research can further refine our understanding of emotional dynamics in consumer behavior, ultimately helping businesses create more meaningful and impactful customer experiences.

Wang, Z., Lao, L., Zhang, X., Li, Y., Zhang, T., & Cui, Z. (2022). Context-dependent emotion recognition. *Journal of Visual Communication and Image Representation*, *89*, 103679.

Wu, L., Mattila, A. S., Wang, C.-Y., & Hanks, L. (2016). The impact of power on service customers' willingness to post online reviews. *Journal of service research*, *19*(2), 224-238.

Xu, W., Yao, Z., He, D., & Cao, L. (2023). Understanding online review helpfulness: a pleasure-arousal-dominance (PAD) model perspective. *Aslib Journal of Information Management*.

Yin, D., Bond, S. D., & Zhang, H. (2014). Anxious or angry? Effects of discrete emotions on the perceived helpfulness of online reviews. *Mis Quarterly*, *38*(2), 539-560.

Yin, D., Bond, S. D., & Zhang, H. (2017). Keep your cool or let it out: Nonlinear effects of expressed arousal on perceptions of consumer reviews. *Journal of marketing research*, *54*(3), 447-463.

You, Y., Vadakkepatt, G. G., & Joshi, A. M. (2015). A meta-analysis of electronic word-of-mouth elasticity. *Journal of marketing*, *79*(2), 19-39.

Zeithaml, V. A. (1988). Consumer perceptions of price, quality, and value: a means-end model and synthesis of evidence. *Journal of marketing*, *52*(3), 2-22.

Zeithaml, V. A., & Parasuraman, A. (2004). Service quality. *Cambridge, MA*.

Zhang, W., Qiu, L., Tang, F., & Sun, H.-J. (2023). Gender differences in cognitive and affective interpersonal emotion regulation in couples: an fNIRS hyperscanning. *Social Cognitive and Affective Neuroscience*, *18*(1), nsad057.

Zhu, F., & Zhang, X. (2010). Impact of online consumer reviews on sales: The moderating role of product and consumer characteristics. *Journal of marketing*, *74*(2), 133-148.
33